\documentclass[12pt]{article}

\usepackage{times}

\usepackage{graphicx}

\newenvironment{sciabstract}{%
\begin{quote} \bf}
{\end{quote}}

\newcounter{lastnote}
\newenvironment{scilastnote}{%
\setcounter{lastnote}{\value{enumiv}}%
\addtocounter{lastnote}{+1}%
\begin{list}%
{\arabic{lastnote}.}
{\setlength{\leftmargin}{.22in}}
{\setlength{\labelsep}{.5em}}}
{\end{list}}

\title{Direct Imaging of Multiple Planets Orbiting the Star HR~8799}

\author
{Christian Marois,$^{1,2,3\ast}$ , Bruce Macintosh,$^{2}$ Travis Barman,$^{4}$\\
 B. Zuckerman,$^{5}$ Inseok Song,$^{6}$ Jennifer Patience,$^{7}$\\
 David Lafreni\`{e}re,$^{8}$ Ren\'{e} Doyon,$^{9}$\\
\\
\normalsize{$^{1}$NRC Herzberg Institute of Astrophysics,}\\
\normalsize{5071 West Saanich Rd, Victoria, BC, V9E 2E7, Canada}\\
\normalsize{$^{2}$Lawrence Livermore National Laboratory, 7000 East Ave, Livermore, CA 94550, USA}\\
\normalsize{$^{3}$Astronomy Department, University of California, Berkeley, CA 94720, USA }\\
\normalsize{$^{4}$Lowell Observatory, 1400 West Mars Hill Road, Flagstaff, AZ 86001, USA}\\
\normalsize{$^{5}$Physics \& Astronomy Department and Center for Astrobiology,}\\
\normalsize{University of California, Los Angeles, CA 90095, USA}\\
\normalsize{$^{6}$}University of Georgia, Physics and Astronomy, 240 Physics, Athens, GA 30602, USA\\
\normalsize{$^{7}$University of Exeter, School of Physics, Stocker Road, Exeter, EX4 4QL, UK}\\
\normalsize{$^{8}$Department of Astronomy and Astrophysics, University of Toronto,}\\
\normalsize{50 St. George Street, Toronto, ON, M5S 3H4, Canada}\\
\normalsize{$^{9}$D\'{e}partement de Physique and Observatoire du Mont M\'{e}gantic, Universit\'{e} de Montr\'{e}al,}\\
\normalsize{C.P. 6128, Succ. Centre-Ville, Montr\'{e}al, QC, H3C 3J7, Canada}\\
\\
\normalsize{$^\ast$To whom correspondence should be addressed; E-mail: christian.marois@nrc-cnrc.gc.ca.}
}

\date{}

\begin{document}


\baselineskip24pt


\maketitle


\begin{sciabstract}
Direct imaging of exoplanetary systems is a powerful technique that can reveal Jupiter-like planets in wide orbits, can enable detailed characterization of planetary atmospheres, and is a key step towards imaging
Earth-like planets. Imaging detections are challenging due to the combined effect of small angular separation and large luminosity contrast between a planet and its host star. High-contrast observations with the Keck and Gemini telescopes have revealed three planets orbiting the star HR 8799, with projected separations of 24, 38, and 68 astronomical units. Multi-epoch data show counter-clockwise orbital motion for all three imaged planets. The low luminosity of the companions and the estimated age of the system imply planetary masses between 5 and 13 times that of Jupiter. This system resembles a scaled-up version of the outer portion of our Solar System.
\end{sciabstract}


During the past decade various planet detection techniques -- precision radial velocities, transits, and microlensing -- have been used to detect a diverse population of exoplanets. However, these methods have two limitations. First, the existence of a planet is inferred through its influence on the star about which it orbits; the planet is not directly discerned [photometric signals from some of the closest-in giant planets have been detected by careful analysis of the variations in the integrated brightness of the system as the planet orbits its star \cite{demming2008}]. Second, these techniques are limited to small (transits) to moderate (precision radial velocity and microlensing) planet-star separation. The effective sensitivities of the latter two techniques diminish rapidly at semi-major axes beyond about 5~AU. Direct observations allow discovery of planets in wider orbits and allow the spectroscopic and photometric characterization of their complex atmospheres to derive their physical characteristics.

There is indirect evidence for planets in orbits beyond 5~AU from their star. Some images of dusty debris disks orbiting main-sequence stars (the Vega phenomenon) show spatial structure on a scale of tens to hundreds of astronomical units \cite{kalas2005}. The most likely explanation of such structure is gravitational perturbations by planets with semi-major axes comparable to the radius of the dusty disks and rings [see references in \cite{zuck2004a}].

The only technique currently available to detect planets with semi-major axes greater than about~5 AU in a reasonable amount of time is infrared imaging of young, nearby stars. The detected near-infrared radiation is escaped internal heat energy from the recently formed planets. During the past decade, hundreds of young stars with ages $\leq $ 100~Ma have been identified within $\sim $100~pc of Earth \cite{zuck2004b,torres2008}, and many of these have been imaged in the near-IR with ground-based adaptive optics (AO) systems and with the Hubble Space Telescope. Direct imaging searches for companions of these stars have detected some objects that are generally considered to be near or above the mass threshold $13.6$~M$_{\rm{Jup}}$ dividing planets from brown dwarfs [see \cite{nakajima1995} for an example and \cite{zuck2008} for a list of known substellar objects orbiting stars], and one planetary mass companion that is orbiting a brown dwarf, not a star \cite{chauvin2004}. Recently, Lafreni\`{e}re et al. \cite{lafreniere2008} have detected a candidate planet near a young (5~Ma) star of the Upper Scorpius association, but a proper motion analysis is required to confirm that it is bound to the host star and not an unrelated low-mass member of the young association. In this issue, Kalas et al. report the detection, in visible light, of a candidate planetary mass companion near the inner edge of the Fomalhaut debris disk \cite{kalas2008}. Non-detections of the candidate companion at near-IR wavelengths suggest that the detected visible flux may be primarily host-star light scattering off circumplanetary dust rather than photons from the underlying object. A statistical Bayesian analysis of a dedicated AO survey of nearby young F-, G- and K-type stars shows that exoplanets are relatively rare at separations $> 20$~AU around stars with masses similar to the Sun \cite{lafreniere2007a}.

Bright A-type stars have been mostly neglected in imaging surveys since the higher stellar luminosity offers a  less favorable planet-to-star contrast. However, main sequence A-type stars do have some advantages. The higher-mass A stars can retain heavier and more extended disks and thus might form massive planets at wide separations, making their planets easier to detect. Millimeter interferometric continuum observations of the nearest Herbig~Ae stars, the precursors to A-type stars, indicate that these are encircled by disks with masses up to several times the Minimum Mass Solar Nebula \cite{mannings1997}, the minimum amount of solar abundance material ($0.01$M$_{\rm{Sun}}$) required to form all planets in the Solar System \cite{Weidenschilling1977}. Associated millimeter line observations resolve these gas disks and indicate that their outer radii are 85-450~AU \cite{mannings1997}. The most exceptional example of a young A-star disk is the one orbiting IRAS 18059-3211, which is estimated to have a mass of 90 times the Minimum Mass Solar Nebula and an outer radius extending to $\sim $3,000~AU \cite{bujarrabal2008}. Radial velocity surveys of evolved A stars do seem to confirm these hypotheses by showing a trend of a higher frequency of planets at wider separations \cite{johnson2007}. In this article, we describe the detection of three faint objects at 0.63$^{\prime \prime}$, 0.95$^{\prime \prime}$ and 1.73$^{\prime \prime}$ (24, 38 and 68 AU projected separation, see Fig.~1) from the dusty and young A-type main sequence star HR~8799, show that all objects are co-moving with HR~8799, and describe their orbital motion and physical characteristics.

\section*{HR~8799 Stellar Properties\label{starhr8799}}

HR 8799 (also V342 Peg \& HIP114189, located 39.4~pc \cite{leeuwen2007} from Earth) is the only star known that has simultaneously been classified as $\gamma$ Doradus (variable), $\lambda$ Bootis (metal-poor Population I A-type star) and Vega-like (far-IR excess emission from circumstellar dust)  \cite{gray1999,sadakane1986}. A fit to the Infrared Astronomical Satellite (IRAS) and Infrared Space Observatory (ISO) photometry indicates that it has a dominant dust disk with temperature of 50~K \cite{zuck2004a,rhee2007}. Such blackbody grains, in an optically thin disk, would reside $\sim $75~AU ($\sim 2^{\prime \prime}$) from HR~8799. This would place the dust just outside the orbit of the most distant companion seen in our images (see Fig. 1), similar to the way the Kuiper Belt is confined by Neptune in our Solar system.

The fractional IR luminosity (L$_{\rm{IR}}$/L$_{\rm{bol}}$ = 2.3$X$10$^{-4}$) \cite{moor2006,rhee2007} is too bright to come from a geometrically thin, flat disk orbiting at such large distances from HR~8799. Such an optically thin disk would need to be warped or Òpuffed upÓ in the vertical direction, plausibly by the gravitational influence of nearby planets. Submillimeter photometry indicates a dust mass of 0.1 Earth masses \cite{williams2006}, making it one of the most massive debris disks detected by IRAS \cite{rhee2007}.

When planets form, gravitational potential energy is released and turned into heat in their interior. As planets do not possess any internal nuclear energy source to maintain their temperature, they cool down and become less luminous with time. For massive planets, this self-luminosity can dominate over their stellar insolation for hundreds of millions or billions of years. With some assumptions on the initial conditions at the time of formation, a planet's mass can be derived simply by estimating the planet's luminosity and the system age. Our age estimate for HR8799 is based on four lines of evidence: the star's galactic space motion, the star's position in a color-magnitude diagram, the typical age of  $\lambda$~Boo and $\gamma$~Dor class stars and the large mass of the HR~8799 debris disk.

Most young stars in the solar neighborhood have Galactic space motions (UVW) that fall in limited ranges.  HR~8799's space motion with respect to the Sun, as calculated from published distance, radial velocity and proper motion, is UVW = (-11.9, -21.0, -6.8 km s$^{-1}$) \cite{leeuwen2007,wilson1953}.  This UVW is similar to that of other stars with an age between that of the TW Hydra association [$8$~Ma \cite{zuck2004b,torres2008}] and the Pleiades [$125 \pm 8$~Ma \cite{stauffer1998}].  The UVW of HR~8799 is similar to that of members of the 30~Ma old, southern hemisphere, Columba and Carina Associations \cite{torres2008}. Calculations of the UVW of the young stars HD~984 and HD~221318, which lie near HR8799, show that their space motions are similar to that of HR8799.  We estimate the ages of HD~984 and HD~221318 to be 30 and 100~Ma, respectively, while the FEPS team estimates the age of HD~984 to be 40~Ma \cite{najita2005}. Overall, the UVW of HR~8799 is clearly consistent with those of young clusters and associations in the solar neighborhood.  Of course, in this UVW range of young stars, there are also older stars with random motions; so other, independent, methods must also be employed to place limits/constraints on the age of HR~8799.

HR 8799 is also found below the main sequence of the Pleiades, $\alpha$ Per (70~Ma) and IC2391 (50~Ma) on a Hertzsprung-Russell diagram. This is consistent with a younger age compared to that of the Pleiades \cite{zuck2001}. Even with the more recent Tycho measurement and correcting for the star's low metallicity, so that B-V is increased and lies between 0.26 and 0.3, HR 8799 still lies low on the Hertzsprung-Russell diagram when plotted against known young stars \cite{zuck2001}, consistent with our young age estimate.

The $\lambda$~Boo stars are generally thought to be young, up to a few 100~Ma \cite{gray2002}. The $\gamma$~Dor class stars are probably also young; they are seen in the Pleiades and in NGC 2516 (age $\sim $100~Ma), but not in the Hyades (age $\sim $650 Ma) \cite{krisciunas1995}.

Finally, the probability that a star has a massive debris disk like HR~8799 declines with age \cite{rhee2007}. Considering all of the above, we arrive at an estimate of 60~Ma and a range between [30,160]~Ma, consistent with an earlier independent estimate of $20-150$~Ma \cite{moor2006}. The conservative age upper limit for HR~8799 is chosen to be the $\sim 5\sigma$ upper limit to the Pleiades age.

\section*{Observations}
The sensitivity of high-contrast ground-based AO imaging is limited primarily by quasi-static speckle artifacts; at large separations ($>0.5^{\prime \prime}$), the main source of speckles is surface errors on the telescope primary mirror and internal optics. To remove this noise, we used angular differential imaging (ADI) in our observations \cite{marois2006,lafreniere2007}. This technique uses the intrinsic field-of-view rotation of altitude/azimuth telescopes to decouple exoplanets from optical artifacts. An ADI sequence is obtained by keeping the telescope pupil fixed on the science camera and allowing the field-of-view to slowly rotate with time around the star. Our observations were obtained in the near-infrared (1.1 to 4.2 microns), a regime where the planets are expected to be bright and where the AO system provides excellent image correction. ADI sequences at various wavelengths were acquired using the adaptive optics system at the Keck and Gemini telescopes and the corresponding facility near-infrared cameras, NIRC2 and NIRI, between 2007 and 2008 (see Fig.~1). During each observing sequence, we typically obtained a mix of unsaturated short-exposure images of the star, to determine its precise location and brightness, together with a set of 30-s exposures that overexposed the central star but had maximum sensitivity to faint companions. Some coronagraphic images were also acquired with NIRC2 to benefit from the simultaneous photometric calibration achievable with a partially transmissive focal plane mask. The b and c companions were first seen in October 2007 Gemini data; the d component was first detected in Keck data in 2008. The b and c components were also visible in a re-analysis of non-ADI Keck data obtained for a related program in 2004 [the data sets and the reduction technique are described further in the supplemental online material (SOM)]

\section*{Astrometric Analysis}

Following the initial detection of the companions, we evaluated their positions relative to the star to confirm that they are co-moving with it (possibly including orbital motion) and not unrelated back/foreground objects (see SOM; Table~1, Table~S2, and Fig.~2). Because HR~8799b was visible in the 2004 Keck images, we have more than 4 years of time baseline for proper motion measurements. With the large proper motion of HR~8799 ($0.13^{\prime \prime}$/year), the HR~8799b object is shown to be bound at a significance of 98 times the estimated 1-sigma uncertainty. Additionally, the data show that it is orbiting counter-clockwise. It moved $25 \pm 2$ mas/year ($0.98$ AU/year) southeast during the 4 year period. Its detected orbital motion is near perpendicular to the line connecting the planet and primary, suggesting that the system is viewed nearly pole-on and that the orbit is not very eccentric. The near face-on perspective is further supported by the slow projected rotational velocity of HR~8799 [$\sim $40 km sec$^{-1}$ \cite{gray1999}]; this is well below average for late-A and early-F type stars \cite{royer2007}. If we assume that it has a semi-major axis of $68$~AU, a circular orbit, a pole-on view, and a host stellar mass equal to $1.5$ Solar masses, then the orbital period and motion of HR~8799b are $\sim $450 years and $0.93$~AU/year ($24$ mas/year) respectively, consistent with our measurements.

HR~8799c is also detected, at lower significance, in the 2004 data set. The measurement of its 4 year proper motion confirms that it is bound to the star at the $90$~sigma level. Its orbit is also counter-clockwise at $30 \pm 2$~mas/year ($1.18$~AU/year). For its semi-major axis of $38$~AU, the orbital period is $\sim 190$ years and the expected orbital motion is $1.25$~AU/year ($32$~mas/year). Again, the orbital motion is close to being perpendicular to the line connecting the planet to the primary.

HR~8799d was first detected in the July 2008 data set. The two months of available proper motion measurements are sufficient to confirm that it is bound to the star at the $\sim $6 sigma level. The available data is also consistent with a counter-clockwise orbital motion of $42 \pm 27$~mas/year ($1.65$~AU/year). For a semi-major axis of $24$~AU, the orbital period is 100 years and the expected orbital motion is $1.57$~AU/year ($40$~mas/year).

\section*{HR~8799bcd Photometric Analysis \label{phot}}

All three companions are intrinsically faint and have red near-IR colors that are comparable to those of substellar-mass objects with low effective temperatures (see Table~1). Compared to old field brown dwarfs (objects with masses between planets and stars), all three companions lie at the base of the L dwarf spectral sequence -- objects known to be cool and have dusty clouds in their atmospheres (Fig.~3).  Two candidate free-floating Pleiades brown dwarfs, with comparable colors and absolute K-band magnitudes to HR~8799c and d, are consistent with a mass of $\sim 11$~M$_{\rm{Jup}}$ from evolutionary models \cite{casewell2007}. If HR~8799 is (as is likely) younger than the Pleiades, the c and d companions would be even less massive . HR~8799b is fainter than all of the known Pleiades substellar members and thus is below $11$~M$_{\rm{Jup}}$ (Fig.~3). All three companions stand apart from the older, more massive brown dwarfs in a color-magnitude diagram. The known distance to HR 8799, and photometry for each companion that covers a substantial fraction of the spectral energy distribution (SED), allow for a robust measurement of the bolometric luminosity (L$_{\rm{bol}}$). We fit a variety of synthetic SEDs (generated with the PHOENIX model atmosphere code) to the observed photometry for each companion assuming their atmospheres were either cloud-free, very cloudy, partly cloudy ($50$\% coverage), or radiated like black bodies.  This fitting process is equivalent to simultaneously determining bolometric corrections for each band-pass for various model assumptions. Luminosities were also obtained using the K-band bolometric corrections for brown dwarfs \cite{golimowski2004}. Although the different models produce different estimates of effective temperature, the range of L$_{\rm{bol}}$ for each object is small (see Table~1), indicating that our estimate is robust against the uncertainty in the details of the atmosphere and clouds (see the SOM for more details).

The cooling of hydrogen-helium brown dwarfs and giant planets is generally well understood; however, the initial conditions associated with the formation of objects from collapsing molecular clouds or core accretion inside a disk are uncertain. Consequently, theoretical cooling tracks of objects at young ages may not be reliable. Recent efforts to establish initial conditions for cooling tracks based on core-accretion models have produced young Jupiter-mass planets substantially fainter ($< 10^{-5}$ L$_{\rm{Sun}}$) than predicted by traditional models \cite{marley2007}. However, these hybrid models do not yet include a realistic treatment of the complex radiative transfer within the accretion-shock and thus provide only lower-limits on the luminosity at young ages. Warmer, more luminous planets originating from core accretion cannot be ruled out.

Although HR~8799 is young, its upper age limit ($\sim 160$~Ma) is near the time when the differences among cooling tracks with various initial conditions are not so dramatic and, given the uncertainties associated with all planet evolution models, standard cooling tracks are as reliable at these ages as other hybrid models. Fig.~4 compares the measured luminosities and age range for HR~8799 bcd to theoretical Òhot startÓ cooling tracks for a variety of masses \cite{baraffe2003}.

The region occupied by all three companions falls below the lowest mass brown dwarf, well inside the planet regime. The masses derived from the luminosities, cooling tracks, and best age for bcd are respectively 7, 10, and 10 M$_{\rm Jup}$.  See table~1 for values of additional important properties derived from the cooling track comparison, with uncertainties based on our current best age range. In the very unlikely event that the star is older than our estimated upper limit, it would need to be $> 300$~Ma for all three objects to be brown dwarfs.

The large planet masses and orbital radii in the HR~8799 system are challenging to explain in the context of a core accretion scenario. A number of factors such as stellar mass \cite{ida2005}, metallicity  \cite{ida2004}, disk surface density \cite{lissauer1987}, and planet migration in the disk \cite{rice2003} influence the core accretion process. The stellar mass of HR~8799 is larger than the Sun. The star's metallicity is low, especially in refractory elements, but for a $\lambda$~Boo star this is usually attributed to the details of the starÕs accretion and atmospheric physics rather than an initial low metallicity for the system \cite{gray2002}.

The exceptionally dusty debris disk around HR~8799 may indicate that the proto-planetary disk was massive and had a high surface density, factors conducive to planet formation. Alternatively, the giant planets in the HR~8799 system may have formed rapidly from a gravitational instability in the early disk \cite{boss1997,rafikov2005}. Some models \cite{rafikov2005} of such instabilities do favor the creation of massive planets ($>6$~M$_{\rm{Jup}}$).

As suggested by the color-magnitude diagram (Fig.~3), each companion appears to be at the edge of (or inside) the transition region from cloudy to cloud-free atmospheres. Current planet atmosphere models have difficulties fitting the color and spectrum features of these objects. The physical mechanism responsible for the clearing of clouds in ultra-cool atmospheres is not fully understood, but recent cloud models with vertical stratification have had some success at simulating/producing photometric properties in this transition region \cite{saumon2008}. A modified PHOENIX atmospheric model was developed that incorporates cloud stratification. These updated models were found to match well-known brown dwarfs located in the cloudy/cloud-free transition region. With the cloud stratification model, PHOENIX is capable of producing spectra that are consistent with the observed photometry and the bulk properties (effective temperature, radius, and gravity) predicted by the cooling tracks (Fig. 5). Clearly these synthetic models do not reproduce all of the photometric data, but given the difficulty of cloud modeling, the agreement is sufficient to support the effective temperatures and radii determined from the cooling tracks.

\section*{Conclusions}

The three co-moving companions of HR~8799 are significantly different from known field objects of similar effective temperature; the only similar object known is the planetary mass companion to the brown dwarf 2M1207. Low luminosities of these companions and the young age for HR~8799 indicate that they have planetary masses and are not brown dwarfs. The nature of the system provides an additional indirect line of evidence for planetary-mass companions (and hence a low age). There are no known systems where multiple brown dwarfs independently orbit an early-type star; the only systems we know of with multiple companions in independent orbits are the exoplanetary systems discovered from the precision radial velocity method. Interestingly, our observations show that the HR 8799 planets orbit in the same direction, similar to the planets in our own solar system and consistent with models of planet formation in a disk. In many ways this resembles a scaled-up version of our solar system. HR~8799 has a luminosity of $4.9$~L$_{\rm{Sun}}$, so the radius corresponding to a given equilibrium temperature is 2.2 times larger than the corresponding radius in our solar system. Because formation processes will be affected by luminosity - e.g. the location of the snow line where water can condense on rocky material to potentially form giant planet cores - one can view the three planetary companions as having temperature-equivalent projected orbital separations of 11, 17 and 31~AU, to be compared with 9.5, 19, and 30~AU for Saturn, Uranus, and Neptune. The HR~8799 planets are also consistent with formation through instabilities in a massive protoplanetary disk, which may form objects with masses above 5~M$_{\rm{Jup}}$ \cite{rafikov2005}, but the core accretion scenario cannot yet be ruled out.

The presence of these massive planets still leaves dynamic room for other Jovian-mass planets or even lower mass terrestrial planets in the inner part of the system. In our survey, we only observed a few early-type stars before making this detection, compared to similar imaging surveys of young G-, K-and M-type stars that have covered more than a few hundred targets. This may indicate that Jovian-mass planetary companions to early-type stars are much more common at separations beyond $\sim $20~AU, consistent with what was suggested by radial velocity surveys of evolved A-type stars \cite{johnson2007}.

\begin{scilastnote}
\item We thank the Keck \& Gemini staff, particularly T. Armandroff, B. Goodrich, and J.-R. Roy for support with the follow-up observations. We thank the UCLA galactic center team, especially J. Lu, for the NIRC2 plate scale and North orientation errors. We are indebted to E. Becklin and R. Racine for their contributions in the earliest stages of this research. CM and DL are supported in part through postdoctoral fellowships from the Fonds Qu\'{e}b\'{e}cois de la Recherche sur la Nature et les Technologies. Portions of this research were performed under the auspices of the US Department of Energy by LLNL under contract DE-AC52-07NA27344, and also supported in part by the NSF Science and Technology CfAO, managed by the UC Santa Cruz under cooperative agreement AST 98-76783. We acknowledge support by NASA grants to UCLA and Lowell Observatory. RD is supported through a grant from the Natural Sciences and Engineering Research Council of Canada. The data were obtained at the W.M. Keck and Gemini Observatories. This publication makes use of data products from the Two Micron All Sky Survey and the SIMBAD database.\\
\end{scilastnote}


\clearpage

\vspace{0.5cm}
\begin{tabular}{lccc}
\multicolumn{4}{c}{\bf Table 1: HR~8799 planetary system data} \\
\hline
\multicolumn{4}{c}{HR~8799}\\ \hline \hline
Spectral type & \multicolumn{3}{c}{A5V} \\
Mass & \multicolumn{3}{c}{$1.5 \pm 0.3$~M$_{\rm{Sun}}$} \cite{gray1999}\\
Luminosity & \multicolumn{3}{c}{$4.92 \pm 0.41$ L$_{\rm{Sun}}$} \cite{gray1999}\\
Distance & \multicolumn{3}{c}{ $39.4 \pm 1.0$ pc ($128 \pm 3$~ly)} \cite{leeuwen2007}\\
Proper motion [E,N] & \multicolumn{3}{c}{ $[107.88 \pm 0.76$,$-50.08 \pm 0.63]$ mas/year} \cite{leeuwen2007}\\
Age & \multicolumn{3}{c}{$60 [30-160]$~Ma} \\
Metallicity & \multicolumn{3}{c}{ $\log([\rm{M/H}]/[\rm{M/H}]_{\rm{Sun}}$) $= -0.47$} \cite{gray1999}\\
J, H, Ks, L' & \multicolumn{3}{c}{$5.383 \pm 0.027$, $5.280 \pm 0.018$, $5.240 \pm 0.018$, $5.220 \pm 0.018$ }\\ \hline
& \multicolumn{3}{c}{Separation w.r.t the host star in [E, N]$^{\prime \prime}$}\\
& HR~8799b & c & d\\
2004 July 14 ($\pm 0.005^{\prime \prime}$) & $[1.471, 0.884]$ & $[-0.739, 0.612]$ & $-$\\
2007 Oct. 25 ($\pm 0.005^{\prime \prime}$) & $[1.512, 0.805] $& $[-0.674, 0.681] $& $-$\\
2008 July 11 ($\pm 0.004^{\prime \prime}$) & $[1.527, 0.799] $& $[-0.658, 0.701] $& $[-0.208, -0.582]$\\
2008 Aug. 12 ($\pm 0.002^{\prime \prime}$)& $[1.527, 0.801] $& $[-0.657, 0.706] $& $[-0.216, -0.582]$\\
2008 Sept. 18 ($\pm 0.003^{\prime \prime}$)& $[1.528, 0.798] $& $[-0.657, 0.706] $& $[-0.216, -0.582]$\\ \hline
Projected Sep. (AU) & $68$ & $38$ & $24$ \\
Orbital Motion ($^{\prime \prime}$/year) & $0.025 \pm 0.002$ & $0.030 \pm 0.002$& $0.042 \pm 0.027$\\
Period for pole view& & & \\
cir. orbits (years) & $\sim 460$ & $\sim 190$& $\sim 100$\\ \hline
M$_{\rm{J}}$ ($1.248 \mu$m) & $16.30 \pm 0.16 $ & $14.65 \pm 0.17$ & $15.26 \pm 0.43$\\
M$_{\rm{H}}$ ($1.633 \mu$m) & $14.87 \pm 0.17$ & $13.93 \pm 0.17$ & $13.86 \pm 0.22$ \\
M$_{\rm{CH4S}}$ ($1.592 \mu$m) & $15.18 \pm 0.17$ & $14.25 \pm 0.19$ & $14.03 \pm 0.30$ \\
M$_{\rm{CH4L}}$ ($1.681 \mu$m) & $14.89 \pm 0.18$ & $13.90 \pm 0.19$ & $14.57 \pm 0.23$ \\
M$_{\rm{Ks}}$ ($2.146 \mu$m) & $14.05 \pm 0.08$ & $13.13 \pm 0.08$ & $13.11 \pm 0.12$ \\
M$_{\rm{L'}}$ ($3.776 \mu$m) & $12.66 \pm 0.11$ & $11.74 \pm 0.09$ & $11.56 \pm 0.16$ \\ \hline
Luminosity (L$_{\rm{Sun}}$) & $-5.1 \pm 0.1$ & $-4.7 \pm 0.1$ & $-4.7 \pm 0.1$\\
T$_{\rm{eff}}$ (K)        & $870 [800-900]$ & $1090 [1000-1100]$ & $1090 [1000-1100]$\\
Radius (R$_{\rm{Jup}}$)     &  $1.2 [1.1-1.3]$ &  $1.2 [1.2-1.3]$ & $1.2 [1.2-1.3]$ \\
Mass (M$_{\rm{Jup}}$)       &  $7 [5-11]$ & $10 [7-13]$  & $10 [7-13]$\\ \hline
\end{tabular}
\vspace{0.5cm}
\clearpage

\begin{figure}[htbp]
\begin{center}
\includegraphics[width=13.5cm]{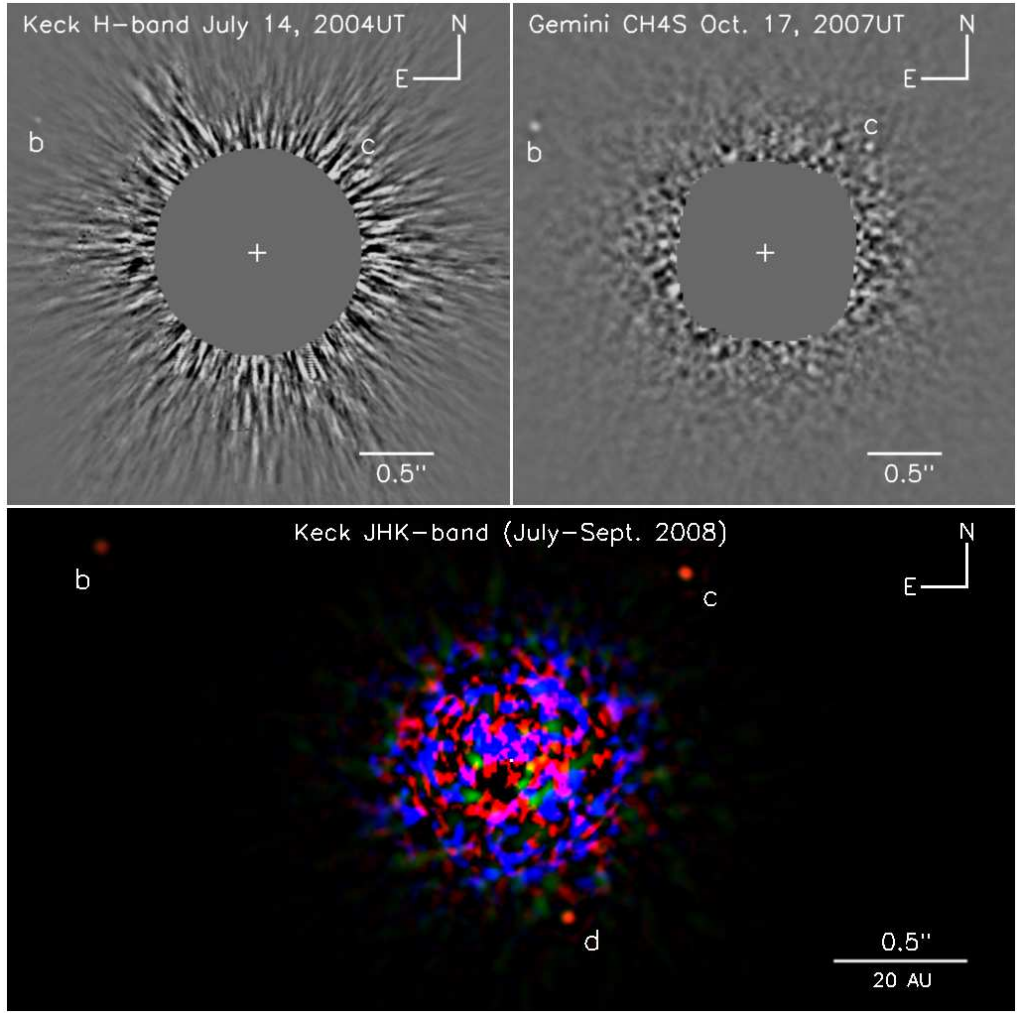}
\end{center}
\caption{HR 8799bcd discovery images after the light from the bright host star has been removed by ADI processing. (Upper left) A Keck image acquired in July 2004. (Upper right) Gemini discovery ADI image acquired in October 2007. Both b and c are detected at the 2 epochs. (Bottom) A color image of the planetary system produced by combining the J-, H-, and Ks-band images obtained at the Keck telescope in July (H) and September (J and Ks) 2008. The inner part of the H-band image has been rotated by 1 degree to compensate for the orbital motion of the d between July and September. The central region is masked out in the upper images but left unmasked in the lower to clearly show the speckle noise level near d.}
\end{figure}
\clearpage

\begin{figure}[htbp]
\begin{center}
\includegraphics[width=\textwidth]{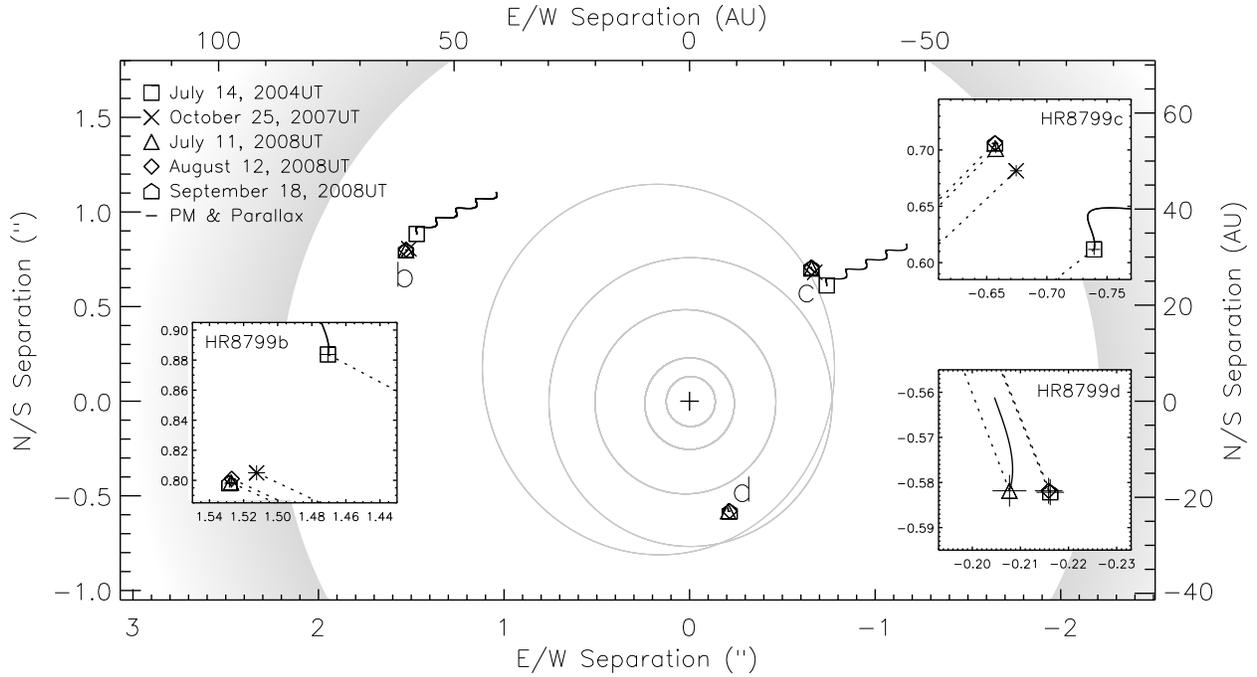}
\end{center}
\caption{HR~8799bcd astrometric analysis. The positions of HR~8799bcd at each epoch are shown in both the overall field of view and in the zoomed-in insets. The solid oscillating line originating from the first detected epoch of each planet is the expected motion of a unbound background objects relative to the star over a duration equal to the maximum interval over which the companions were detected (4 years for b and c, two months for d.) All three companions are confirmed as co-moving with HR~8799 to 98$\sigma $ for b, 90$\sigma $ for c and $\sim 6\sigma $ for d. Counter-clockwise orbital motion is observed for all three companions. The dashed lines in the small insets connect the position of the planet at each epoch with the star. A schematic dust disk -- at 87 AU separation to be in 3:2 resonance with b while also entirely consistent with the far-infrared dust spectrum -- is also shown. The inner gray ellipses are the outer Jovian-mass planets of our Solar system (Jupiter, Saturn, Uranus \& Neptune) and Pluto shown to scale.}
\end{figure}
\clearpage

\begin{figure}[htbp]
\begin{center}
\includegraphics[width=\textwidth]{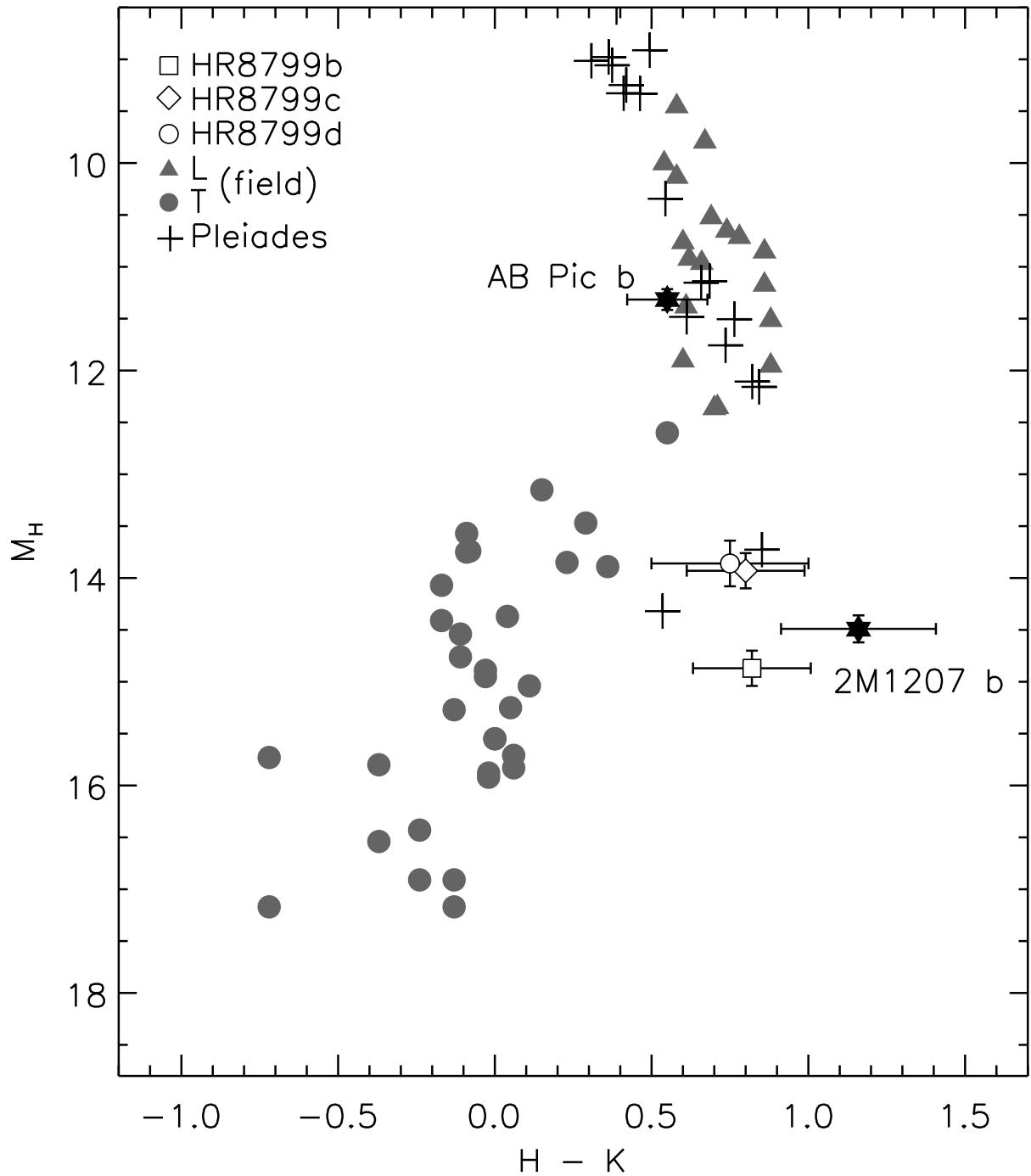}
\end{center}
\caption{Absolute magnitude in H-band versus H-K color. Old, field (grey dots) and young Pleiades brown dwarfs (plusses) are shown along with 2 very low-mass brown dwarfs/planetary mass companions (filled black symbols). Open symbols are HR~8799b (square), c (diamond), and d (circle).}
\end{figure}
\clearpage

\begin{figure}[htbp]
\begin{center}
\includegraphics[width=14cm]{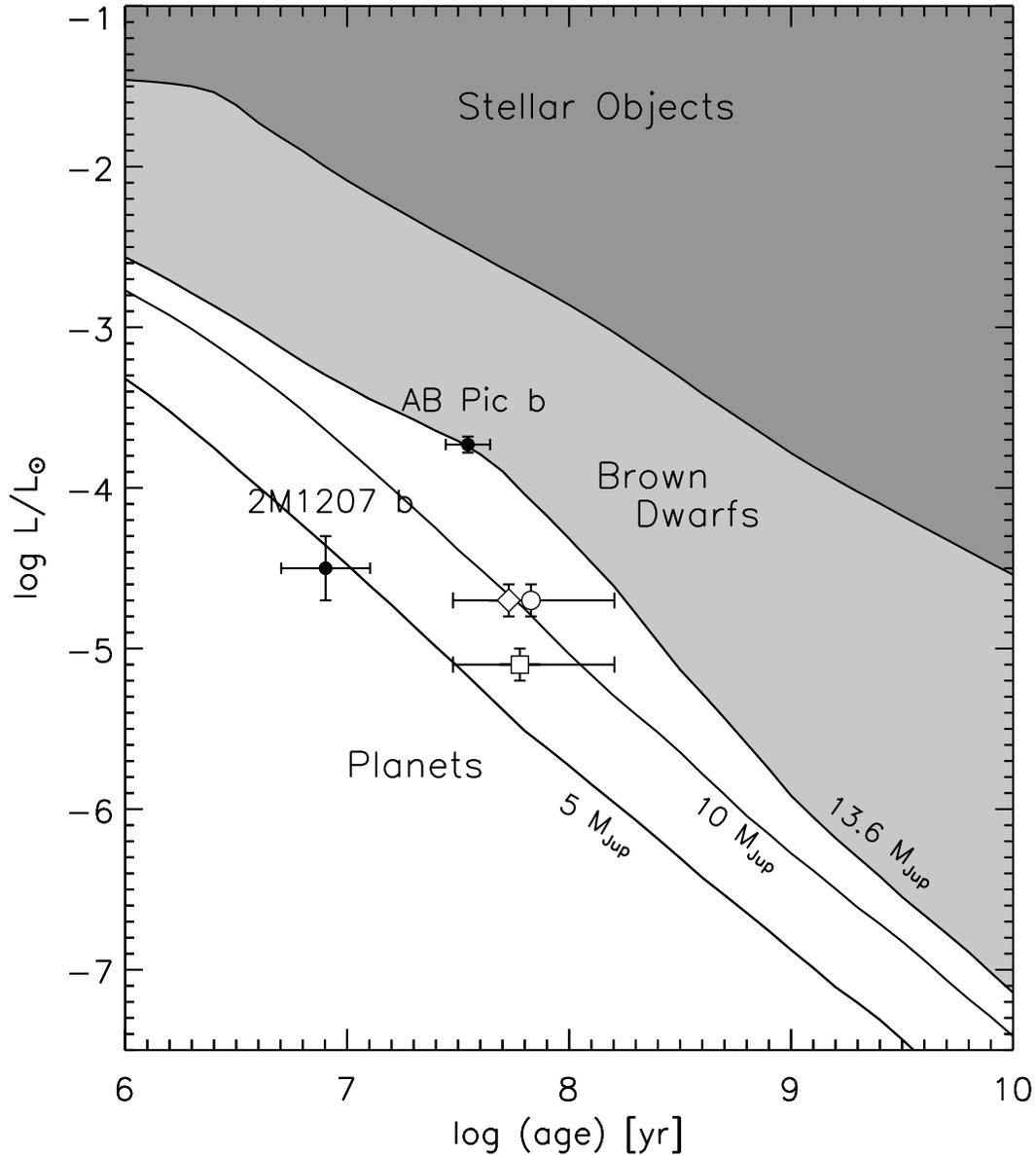}
\end{center}
\caption{Luminosity versus time for a variety of masses \cite{baraffe2003}.  The three coeval points are HR~8799b (square), c (diamond), and d (circle); c and d data points are displaced horizontally for clarity. The locations of the low mass object AB Pic b on the planet/brown dwarf dividing line and a planetary mass companion (2M1207b) to the brown dwarf 2M1207 are also shown (note that alternative models proposed for 2M1207 lead to somewhat larger luminosity and mass ($\sim $8 M$_{\rm{Jup}}$) for the companion \cite{ducourant2008}). The deuterium burning mass limit, currently believed to be $\sim $13.6~M$_{\rm{Jup}}$, has been incorporated into a ``working definition'' of a planet by the International Astronomical Union and is used here to separate planets (which also must orbit a star) from brown dwarfs. The boundary between stars and brown dwarfs is set by stable hydrogen burning.}
\end{figure}
\clearpage

\begin{figure}[htbp]
\begin{center}
\includegraphics[width=\textwidth]{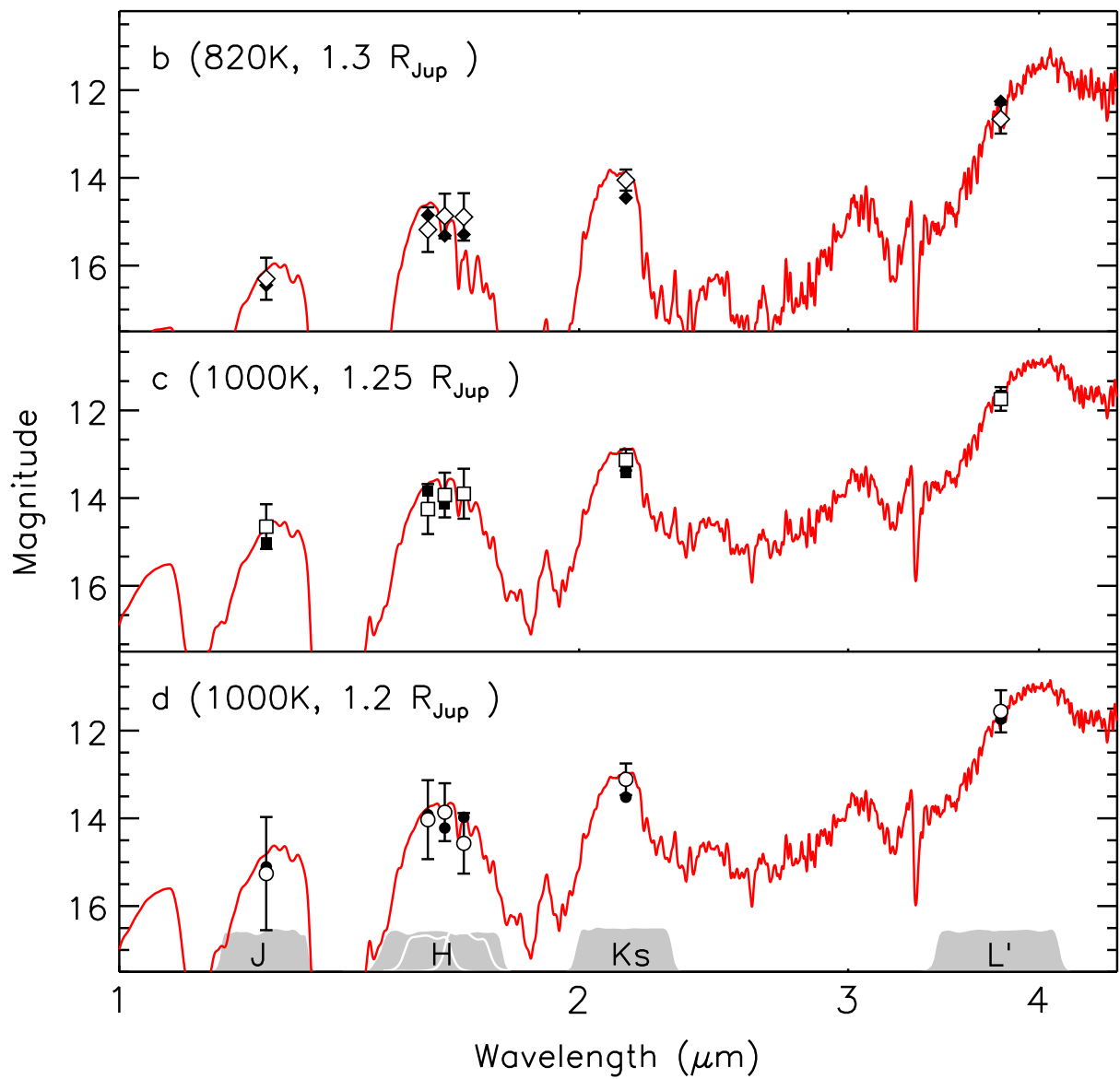}
\end{center}
\caption{Synthetic spectra from model atmospheres containing clouds located between 10 and 0.1 bar of pressure are compared to the measured fluxes (with 3~sigma error bars) for HR~8799 b, c and d. Response curves for each filter band pass are indicated along the x-axis. The predicted magnitudes from the synthetic spectra, averaged over the filter passbands, are shown by the filled symbols.}
\end{figure}
\clearpage

\section*{Supplemental Online Material}

\subsection*{ADI Observations and Data Reduction}
The first ADI imaging sequence on HR~8799 was obtained on October 17th, 2007 at the Gemini Observatory (program GN-2007B-Q-77) using the ALTAIR AO system ({\it S1}) and the NIRI camera ({\it S2}). After the first set of observations, follow-up multi-band imaging was carried out at the Gemini telescope (program GN-2008B-Q-64) and the W.M. Keck II telescope with its AO system ({\it S3}) along with the NIRC2 narrow-field camera ({\it S4}). To derive accurate bolometric luminosities, we have acquired a set of J-, H-, K'-, Ks- and L'-band images that span most of the spectral energy output of the companions. In addition, we have acquired a set of CH4S- and CH4L-band data (two filters that are in and out of the $1.6 \mu$m methane absorption feature seen in cloud-free cool atmospheres) to search for a methane spectral signature. All the data were acquired in the linear regime ($<$ 4\%) of the detector. Once the planetary companions were noted in the 2007-2008 data we re-analyzed non-ADI NIRC2 data obtained on July 10 2004 at the Keck Observatory for a related project, and detected the two outermost companions of the system. In all data sets, both the 38 and 68~AU companions are detected. The 24~AU companion is only detected in the most sensitive ADI image sets.

The data reduction for all ADI sequences are performed using a custom IDL script following a standard reduction technique ({\it S5}). A shutter-closed dark image having the same observing parameters is first subtracted to remove the detector electronic bias and the image is divided by a flat field (produced by uniform illumination) to normalize the spatial variations in sensitivity. For the longer-wavelength K', Ks and L' images, we subtract a blank-sky image to remove thermal backgrounds. Deviant pixels are then removed by interpolating adjacent pixels and image distortions are corrected (for Gemini, the IDL observatory script is used, while for Keck the IDL script {\it nirc2warp.pro} is applied). We coarsely register the images using the unsaturated data and a cross-correlation technique is used for fine registration of all the saturated/occulted images to sub-pixel accuracy. A $21 \times 21$ pixel unsharp mask filter is then applied to remove the low-spatial frequency noise (this filter is applied to both the unsaturated and saturated data). The improved ADI LOCI data reduction scheme ({\it S6}) is used for each individual image to select a set of reference images in which the stellar PSF is sufficiently similar but the field-of-view has rotated enough to not attenuate the companions. After subtracting the reference from each frame we de-rotate and average the frames to produce the final combined image. Fig.~1 shows the resulting data reduction for a subset of the HR~8799 acquired data.

The exoplanet relative positions are obtained at each epoch by fitting a Gaussian function to the planet intensity profile. If multiple data sets are available for a given epoch, the position of the exoplanets are averaged together to reduce random errors. The measured positions were then compared with the expected drift for an unbound background/foreground object from HR8799's space motion and the Earth-induced parallax effect. Due to the large number of data sets available, the higher signal-to-noise ratio of individual sets and well characterized astrometry, only the Keck data has been considered to perform the proper motion analysis. The relative detector positions of the exoplanets and the star are transformed to sky coordinates using the NIRC2 narrow camera plate scale value ($9.963 \pm 0.005$~mas/pixel) and North orientation [$0.13 \pm 0.02$ degree, {\it S7}]. Due to the uncertainty in registering the saturated images at the image center to perform the ADI processing and field rotation alignment, a conservative 0.5 pixel ($\sim $5~mas) 1 sigma centroid uncertainty is added for each data set at each epoch. Note that uncertainty in both the plate scale and North orientation are well below the calculated centroid accuracy of our companions. Since all astrometric observations were acquired using NIRC2, whose plate scale is known to be very stable with time [$<$0.02\% variations of plate scale since 2004, {\it S7}], the absolute plate scale and orientation errors are not included in the positional uncertainties given here. If the Keck astrometry is compared to that from other telescopes, those uncertainties would need to be included. Table~S2 summarizes HR~8799bcd astrometry for all acquired data sets.

Accurate exoplanet photometry is essential to derive the companions' basic physical characteristics. Due again to higher SNRs and the fact that d is detected at almost all wavelenghs, the Keck data is preferred for the photometry analysis. For J-, Ks- and L'-band, we have chosen the coronagraphic data over the earlier saturated PSF sequences due to the better photometric calibration that it provides. The exoplanets' fluxes are derived after convolving the combined ADI-reduced images by a circular aperture having a diameter equal to the wavelength-dependent full-width-at-half-maximum of the telescope diffraction limit. The relative star-to-planet relative intensity is obtained by taking the ratio of the peak fluxes (after aperture convolution) of the average unsaturated stellar images to the planetary fluxes. For coronagraphic long exposure data, the relative planet-star photometry is obtained from the unsaturated PSF core detected through the partially transmissive focal plane mask. The coronagraph throughput has been calibrated using unsaturated unocculted images. The ADI processing does somewhat attenuate the fluxes from the companions. The stellar PSF obtained from the unsaturated data was used to introduce artificial sources into the raw images at various separations, position angles and at intensities similar to the ones of the detected exoplanets.  These artificial images were processed using the same pipeline to calibrate the ADI algorithm throughput as a function of position (the Fig.~1 detection image has been renormalized to show a throughput of 1 at all separations). Individual magnitudes for each companion were derived by calculating the relative intensity (for each bandpass) of the planets compared to that of the host star. The star infrared magnitudes were taken from 2MASS and its L' magnitude was derived from {\it S8}.  The derived photometry (absolute magnitudes) at all epochs and wavelengths can be found in Table~1. Photometric errors are estimated from several sources. First, the variations in peak intensity of successive images of the unsaturated PSF (typically $\sim 5$\%). Second, the speckle and photon noise in the scattered starlight halo, estimated from image statistics at the separation of each companions (data set and separation dependent, from 1\% to 43\%). Third, the ADI-induced photometric error from possible registration offsets from the exact rotational center of the field of view ($\sim 0.5$ pixel RMS resulting in a 3\% error). Fourth, the residual registration accuracy between images ($\sim 0.5$ pixels inducing a 3\% error) and finally the ADI flux renormalization error (typically $\sim 5$\%) estimated from the artificial sources. For non-coronagraphic ADI sequences, an additional $15$\% photometric error is added to account for potential seeing variations during long saturated image sequences (this fluctuation was estimated from coronagraphic data).

\subsection*{Planet Atmosphere Modeling}

To determine the luminosity of each planet, synthetic spectra from a variety of planetary atmosphere models were fit to the photometric measurements described above. These models and spectra were generated using the {\tt PHOENIX} atmosphere code ({\it S9}) which includes two methods for describing the impact of cloud formation from solid (and liquid) material that condenses under the assumption of chemical equilibrium. These two cloud models cover the cases of extremely cloudy, with nearly uniform coverage both vertically and horizontally in the atmospheres, and completely cloud-free ({\it S10}). It is known from studies of brown dwarfs (which experience similar atmospheric conditions as giant planets) that atmospheres transition from having very cloudy atmospheres to nearly cloud-free atmospheres as they cool (called the L-T transition).  Consequently, cloud models that are intermediate between the two extreme cases are required to reproduce the diversity observed in ultra-cool objects.  Since the planets orbiting HR~8799 have photometric properties similar to brown dwarfs at or near the L-T transition, neither of the two extreme cloud models are likely to be appropriate. To better describe the planets, an intermediate cloud model similar to that described by {\it S11} was added to {\tt PHOENIX}. In this case, the clouds are more confined vertically in the atmosphere than in the extreme DUSTY case, where the number density of cloud particles decreases with increasing height. While this cloud model is certainly more realistic than the extreme cloudy assumption, modeling clouds in giant planet atmospheres is very difficult and a wide variety of theoretical attempts are actively being explored and, so far, no clear physical picture has emerged ({\it S12}).

The synthetic spectra from these models were convolved with the response curves of each near-IR filter (Fig.~5, lower panel) used for the observations, thereby generating synthetic photometry.  A standard Levenberg-Marquardt least-squares minimization procedure was used to fit these synthetic photometry to the real data.  As already mentioned in the main article, the best fit temperatures and radii for cloud assumptions were very different, while the luminosities were very similar. This is to be expected since the observations cover a large fraction of the total spectral energy distribution.  As a further check on the luminosity, black bodies and semi-empirical bolometric corrections were used and resulted in nearly identical values (see Table~1 for the luminosities and uncertainties).

Interestingly, the best fit effective temperatures from the model atmosphere analysis are high (ranging from 1700 to 1400K). Such high effective temperatures would imply either unrealistically small radii (a few tenths the size of Jupiter) or that each planet has its own disk exactly aligned edge-on with our line-of-sight causing several magnitudes of extinction. This later situation is very unlikely since edge-on disks are rare, and the observed orbital motions along with the small projected rotation of the host star suggest that HR~8799 is viewed closer to face-on.  The most likely explanation for the unreasonably high effective temperatures implied by the best-fit atmosphere models is missing atmospheric physics (e.g., poor cloud modeling and possibly non-equilibrium chemistry).

Since the theoretical cooling tracks are far less sensitive to the assumptions regarding atmospheric clouds, the masses, effective temperatures, and radii for each planet were obtained from cooling tracks (see Fig.~4) rather than the best fit values to the synthetic photometry mentioned above. However, as a consistency check, synthetic planet spectra were computed using our intermediate cloud model and the parameters determined from the cooling tracks.  These synthetic spectra are compared to the observed data in Fig.~5 and while they do not represent a best-fit to the data, they are a reasonably close match and lend confidence to the values implied by the luminosity and cooling track comparison.  Some of the discrepancies with these synthetic spectra could be alleviated by including non-equilibrium chemistry which has been shown to impact the strength of CH$_4$ and CO bands ({\it S13}).

\clearpage
\subsection*{SOM Tables}

\vspace{0.5cm}
\begin{tabular}{lccc}
\multicolumn{4}{c}{\bf Table S1: HR~8799 observing log} \\
\hline
Inst./Telescope & UT Date & Filter & Saturated Data \\
 & & & Total Exp. Time (s)\\ \hline \hline
NIRC2/Keck & 2004 July 14 & H & $480$ \\
& 2007 Oct. 25 & CH4S & $2700 $\\
& 2008 July 11 & H & $1740$ \\
& & CH4S & $1680$ \\
& 2008 Aug. 12 & J & $900$\\
& & H & $420$\\
& & CH4L & $870$\\
& & K' & $840 $\\
& & L' & $900$ \\
& 2008 Sept. 18 & J &$ 1370$\\
& & Ks &  $1200 $\\
& & L' & $2800$ \\
NIRI/Gemini & 2007 Oct. 17 & CH4S & $3600$\\
\hline
\end{tabular}
\vspace{0.5cm}

\clearpage
\vspace{0.5cm}
\begin{tabular}{lcccc}
\multicolumn{5}{c}{\bf Table S2: NIRC2 HR~8799bcd astrometry.}\\
\hline
UT Date & Filter & \multicolumn{3}{c}{Separation w.r.t the host star in [E, N]$^{\prime \prime}$}\\
 & & b & c & d\\ \hline \hline
2004 July 14 & H & $[1.471, 0.884]$ & $[-0.739, 0.612]$ & -\\
2007 Oct. 25 & CH4S & $[1.512, 0.805]$ & $[-0.674, 0.681]$ & -\\
2008 July 11 & H & $[1.528, 0.805]$ & $[-0.656, 0.703]$ & $[-0.204, -0.579]$\\
& CH4S & $[1.526, 0.792]$ & $[-0.660, 0.698]$ & $[-0.211, -0.585]$\\
Avg. ($\pm 0.004^{\prime \prime}$) & & $[1.527, 0.799]$ & $[-0.658, 0.701]$ & $[-0.208, -0.582]$\\
2008 Aug. 12 & J & $[1.526, 0.801]$ & $[-0.659, 0.702]$ & -\\
& H        & $[1.527, 0.803]$ & $[-0.657, 0.708]$ & -\\ 
& CH4L & $[1.522, 0.797]$ & $[-0.662, 0.703]$ & $[-0.217, -0.582]$\\
& K' & $[1.527, 0.802]$ & $[-0.657, 0.703]$& $[-0.213, -0.580]$\\
& L' & $[1.531, 0.802]$ & $[-0.652, 0.713]$ & $[-0.217, -0.584] $\\
Avg. ($\pm 0.002^{\prime \prime}$) & & $[1.527, 0.801]$ & $[-0.657, 0.706]$ & $[-0.216, -0.582]$ \\
2008 Sept. 18 & J & $[1.527, 0.796]$ & $[-0.656, 0.705]$ & $[-0.213,-0.582]$\\
& Ks & $[1.525, 0.796]$ & $[-0.659, 0.708]$& $[-0.221, -0.581]$\\
& L' & $[1.532, 0.802] $& $[-0.657, 0.703]$ & $[-0.214, -0.584]$ \\
Avg. ($\pm 0.003^{\prime \prime}$) & & $[1.528, 0.798]$ & $[-0.657, 0.706]$ & $[-0.216, -0.582]$ \\
\hline
\end{tabular}
\vspace{0.5cm}

\clearpage
\subsection*{SOM References}
\begin{enumerate}

\item[S1] G. Herriot et al., {\it Proc. SPIE} {\bf 3353}, 488 (1998).
\item[S2] K. W. Hodapp et al., {\it PASP} {\bf 115}, 1388 (2003).
\item[S3] P. Wizinowich et al., {\it PASP} {\bf 112}, 315 (2000).
\item[S4] I. S. McLean \& F.~H. Chaffee, {\it Proc. SPIE} {\bf 4008}, 2 (2000).
\item[S5] C. Marois et al., {\it Astrophys J.} {\bf 641}, 556 (2006).
\item[S6] D. Lafreni\`{e}re et al., {\it Astrophys J.} {\bf 660}, 770 (2007).
\item[S7] A. M. Ghez et al., {\it Astrophys J.}, arXiv:0808.2870 (2008).
\item[S8] A. N. Cox, Allen's Astrophysical Quantities, 4th ed. 2000. 2nd printing, (2001).
\item[S9] P. H. Hauschildt, F. Allard, E. Baron, {\it Astrophys J.} {\bf 512}, 377 (1999).
\item[S10] F. Allard et al., {\it Astrophys J.} {\bf 556}, 357 (2001).
\item[S11] A. S. Ackerman, M.~S. Marley, {\it Astrophys J.} {\bf 556} (2001).
\item[S12] C. Helling et al., {\it Mon. Not. R. Astron. Soc}, arXiv:0809.3657 (2008).
\item[S13] D. Saumon et al., {\it Astrophys J.} {\bf 647}, 552 (2006).

\end{enumerate}

\end{document}